\documentclass[twocolumn,showpacs,preprintnumbers,amsmath,amssymb,pra, superscriptaddress]{revtex4}

\usepackage{graphicx}
\usepackage{dcolumn}
\usepackage{bm}

\def\adot{\dot{\alpha}}

\begin{document}


\title{Collisional perturbation of radio-frequency E1 transitions\\
    in an atomic beam of dysprosium}

\author{A. Cing\"{o}z}
 \email{acingoz@berkeley.edu}
 \affiliation{Department of Physics, University of California at
Berkeley, Berkeley, California 94720-7300}
\author{A.-T. Nguyen}
 \affiliation{Department of Physics, University of California at
Berkeley, Berkeley, California 94720-7300}
\affiliation{University of California, Los Alamos National Laboratory, \\
   Physics Division, P-23, MS-H803, Los Alamos, New Mexico 87545}

\author{D. Budker}
 \email{budker@berkeley.edu}
 \affiliation{Department of Physics, University of California at
Berkeley, Berkeley, California 94720-7300}
 \affiliation{Nuclear Science Division, Lawrence
 Berkeley National Laboratory, Berkeley, California 94720}

\author{S. K. Lamoreaux}
\affiliation{University of California, Los Alamos National Laboratory, \\
   Physics Division, P-23, MS-H803, Los Alamos, New Mexico 87545}
\author{J. R. Torgerson}
  \affiliation{University of California, Los Alamos National Laboratory, \\
   Physics Division, P-23, MS-H803, Los Alamos, New Mexico 87545}

\date{\today}
\begin{abstract}
We have studied collisional perturbations of radio-frequency (rf)
electric-dipole (E1) transitions between the nearly degenerate
opposite-parity levels in atomic dysprosium (Dy) in the presence of
10 to 80 $\mu$Torr of H$_\text{2}$, N$_\text{2}$, He, Ar, Ne, Kr,
and Xe. Collisional broadening and shift of the resonance, as well
as the attenuation of the signal amplitude are observed to be
proportional to the foreign-gas density with the exception of H$_2$
and Ne, for which no shifts were observed. Corresponding rates and
cross sections are presented. In addition, rates and cross sections
for O$_2$ are extracted from measurements using air as foreign gas.
The primary motivation for this study is the need for accurate
determination of the shift rates, which are needed in a laboratory
search for the temporal variation of the fine-structure constant [A.
T. Nguyen, D. Budker, S. K. Lamoreaux, and J. R. Torgerson, Phys.
Rev. A \textbf{69}, 22105 (2004)].
\end{abstract}

\maketitle

\newcommand{\avg}[1]
    {<\!\!#1\!\!>}
\section{Introduction}

\begin{figure*}
\includegraphics{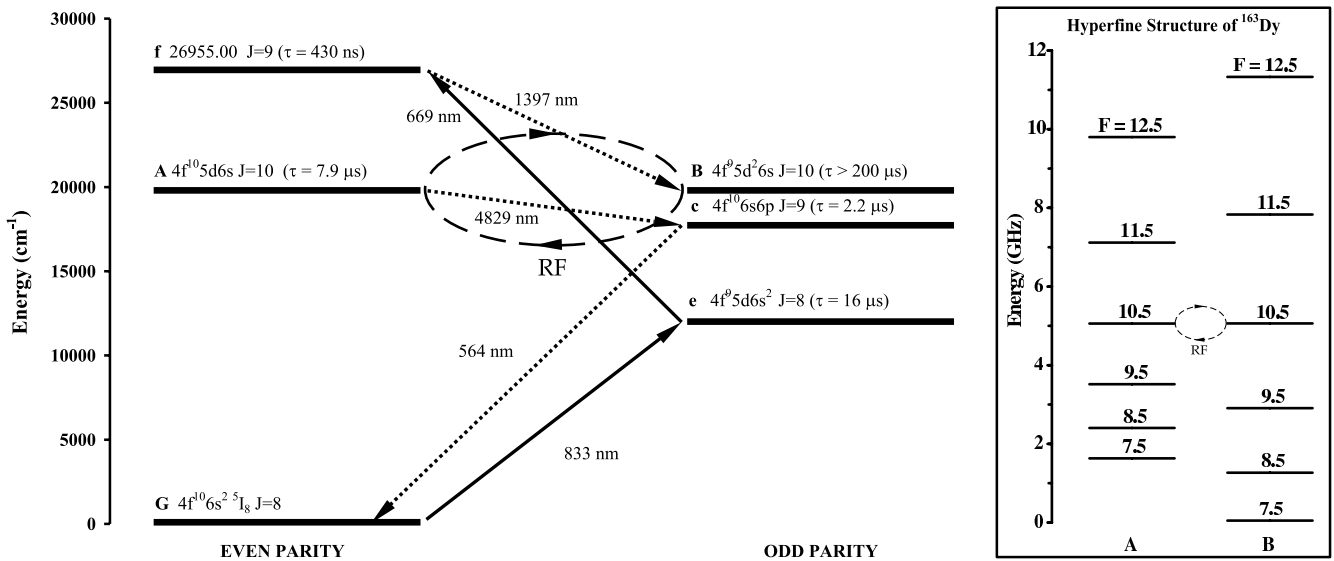} \caption{\label{fig:popsch}
Relevant levels and transitions in atomic dysprosium. Population and
detection scheme: Level B is populated in a three-step process. The
atoms are excited to level f in the first two steps using 833- and
699-nm laser light (solid arrows). The third step is a spontaneous
decay (labeled by a short-dashed arrow) from level f to level B with
$\sim 30$~\% branching ratio. The population is transferred to level
A by the rf electric field (curved long-dashed lines). Atoms in
level A decay to level c and then to the ground state. The
fluorescence from the second decay (564 nm) is used for detection.
Inset: Hyperfine structure of levels A and B for $^{163}$Dy. Zero
energy is chosen to coincide with the lowest hyperfine component.}
\end{figure*}

Perturbations of atomic resonance lines by collisions with atoms
were first observed by Michelson~\cite{Michelson}. With the impact
approximation, Lorentz~\cite{Lorentz} and Weisskopf~\cite{Weisskopf}
provided the earliest theoretical framework. While Lorentz treated
the effects of collisions as a complete termination of a radiation
wave train emitted by the atom, Weisskopf realized that a disruption
in phase was sufficient to explain broadening. However, neither
approach was able to explain shifts of the atomic resonances.
Lindholm~\cite{Lindholm} and Foley~\cite{Foley} remedied this
shortcoming with the adiabatic impact approximation theory which
takes into account the effects of distant collisions that induce
small phase changes ignored by Weisskopf. While this theory is able
to predict the line shape near the resonance, the conditions
necessary for its validity break down near the wings or at higher
pressures where satellite lines and asymmetries have been observed.
The initial theory for satellite lines, called the quasistatic
approach, was developed by Khun~\cite{Khun}. In this approach, the
motion of the perturber is ignored. The line shape is obtained by
averaging the change in the energy difference between the atomic
levels due to the presence of the perturber over the probability
distribution of the stationary perturbers. However, this theory
breaks down for frequencies near resonance or at low pressures.
Since then, these failures have been alleviated by full
quantum-mechanical treatments, developed to unify these two
approaches. These techniques and related experiments are reviewed,
for example, by Ch'en and Takeo~\cite{Chen&Takeo}, Allard and
Kielkopf~\cite{Allard&Kielkopf}, and Szudy and
Baylis~\cite{Szudy&Baylis}.

In this study, we investigate collisional perturbations of
radio-frequency transitions between the nearly degenerate opposite
parity levels of $^{162}$Dy and $^{163}$Dy (Z=66) due to the
introduction of H$_\text{2}$, N$_\text{2}$, He, Ar, Ne, Kr, and Xe
at pressure ranging from 10 to 80 $\mu$Torr. While broadening and
shift rates, and corresponding cross sections are presented for
$^{162}$Dy, only shift results are presented for $^{163}$Dy due to
large statistical uncertainties resulting from the dilution of the
level population by hyperfine levels. Rates and cross sections due
to O$_2$ are extracted from measurements using air as foreign gas.
In addition, upper bounds on collisional-quenching cross sections
are extracted from observations of the signal amplitude as a
function of foreign-gas pressure.

The system under consideration is unusual for a collisional study
since the foreign-gas pressures are $\lesssim~80$~$\mu$Torr. In this
pressure regime, the time between collisions is longer than the
total transit time of the Dy atoms across the apparatus, assuming a
typical cross section of 10$^{-14}$~cm$^2$. The only other study of
collisional perturbations in Dy that we are aware of was performed
with a magnetic trap where the spin relaxation of the ground-state
Zeeman sublevels due to He was presented~\cite{Hancox}.

The primary motivation for this study is the need for accurate
determination of the shift rates which are important for a search
for the temporal variation of the fine-structure constant,
$\alpha$~\cite{Nguyen2004}. Recently, evidence for variation in
$\alpha$ on cosmological time scales was discovered in quasar
absorption spectra~\cite{Webb, Murphy}, corresponding to $\adot
/\alpha = (6.40 \pm 1.35)\times 10^{-16}$/yr assuming a linear shift
over $10^{10}$~years~\cite{Murphy}. The current terrestrial evidence
comes from an analysis of fission products of a natural reactor in
Oklo (Gabon) that operated $1.8 \times 10^9$~years ago,
corresponding to $\adot /\alpha = (-2.3^{+0.8}_{-0.4})\times
10^{-17}$/yr~\cite{Lamoreaux}. However, observational measurements
of this kind are difficult to interpret due to assumptions and
systematic uncertainties. In fact, both the astrophysical evidence
and the Oklo analysis are in disagreement with different quasar
absorption measurements~\cite{Quast, Srianand} and earlier Oklo
analyses~\cite{Damour, Fujii}, respectively. These results have
sparked interest in laboratory searches (see, for example, Refs.
\cite{Marion, Bize, fischer, peik}), which are easier to interpret
but place weaker limits. The best laboratory limit of $|\adot
/\alpha|<1.2\times 10^{-15}$/yr was obtained from a comparison of a
Hg$^+$ optical clock to a Cs microwave clock~\cite{Bize}.

Recent calculations~\cite{Dzuba} have shown that the nearly
degenerate opposite-parity levels in atomic dysprosium
(Fig.~\ref{fig:popsch}, levels A and B) are highly sensitive to
variations in $\alpha$. In Ref.~\cite{Nguyen2004}, a possible
ultimate experimental sensitivity of $|\adot /\alpha|\sim
10^{-18}$/yr was estimated for this system from an analysis of
statistical and systematic uncertainties. In order to realize this
level of sensitivity, the transition frequencies must be kept stable
to a mHz level over a period of a year. Thus, knowledge of
collisional shift rates is critical for the determination of the
maximum allowable pressure in the vacuum chamber and, if necessary,
for correction of the frequency measurements.

In Section II, we describe the technique used to measure transition
frequencies and the procedure used during collisional-perturbation
measurements. Section III presents the results and the details of
the analysis. Section IV considers various systematic uncertainties.
In Section V, we extract broadening and shift cross sections. In
Section VI we present the implications for the $\alpha$-variation
experiment.

\section{Experimental Technique}
\subsection{Overview}
The energy difference between the nearly degenerate levels
considered in Refs.~\cite{Dzuba} and~\cite{Nguyen2004} are on the
order of hyperfine splittings (Fig.~\ref{fig:popsch} inset) and
isotope shifts. Since Dy has seven stable isotopes, there are many
rf transitions. The 235-MHz $^{162}$Dy (4f$^9$5d$^2$6s J=10
$\rightarrow$ 4f$^{10}$5d6s J=10) transition and the 3.1-MHz (F=10.5
$\rightarrow$ F=10.5) transition are used for the $\alpha$-variation
experiment due to practical considerations such as high counting
rate and low frequency, respectively. As shown in
Fig.~\ref{fig:popsch}, the population transfer from the ground state
to the long-lived ($\tau_B >$ 200 $\mu$s~\cite{Budker94}) odd-parity
level B requires three transitions. The first two transitions are
induced via 833- and 669-nm laser light, while the last transition
is a spontaneous decay with a $\sim$ 30~\% branching
ratio~\cite{Nguyen1997}. Atoms are transferred to the even-parity
level A ($\tau_A$=7.9 $\mu$s~\cite{Budker94}) with a
frequency-modulated rf electric field referenced to a commercial Cs
frequency standard. The transition frequency is determined via a
lock-in technique by monitoring the 564-nm fluorescence.

\subsection{Apparatus}
\begin{figure}
\includegraphics{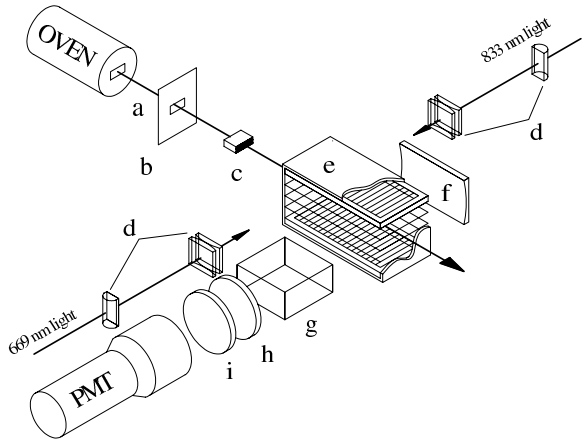}
\caption{\label{fig:oven&int} Experimental setup (not shown to
scale): a) Atomic beam produced by effusive oven source at $\sim
1500$ K; b) atomic-beam collimator; c) oven light collimator; d)
cylindrical lenses to diverge laser beams; e) interaction region of
atoms with the electric field enclosed in a magnetic shield (not
shown); f) cylindrical mirror to collect fluorescent light; g)
lucite light pipe; h) interference filter; and i) short-pass
filter.}
\end{figure}

In this section, we describe the atomic-beam apparatus, the optical
setup, and the detection and pressure-measurement systems.

A detailed description of the atomic-beam source is given in
Ref.~\cite{Nguyen1997}. The beam is produced by an effusive oven
with a multislit nozzle-array operating at $\sim$~1500~K. The oven
consists of a molybdenum tube containing dysprosium metal,
surrounded by resistive heaters made from tantalum wire inside
alumina ceramic tubes. The radiation shielding is provided by five
layers of tantalum sheets surrounding the oven and the heaters. In
addition to the multislit nozzle, two external collimators are used
to collimate the atomic beam and the oven light. The latter is
necessary to minimize the background due to scattered oven light.
The resulting atomic beam has a mean velocity of $\sim 5\times 10^4$
cm/s with a full-angle divergence of $\sim 0.2$ rad (1/$e^2$ level)
in both transverse directions.

The atoms enter the interaction region (Fig.~\ref{fig:oven&int}),
where they are first transferred to level B by two laser-induced
transitions and a spontaneous decay. Approximately 600~mW of 833-nm
light is produced by a Ti:Sapphire ring laser (Coherent 899) pumped
by 15~W of Ar ion laser light (Coherent Innova 400). A ring dye
laser (Coherent 699) with DCM
[4-(Dicyanomethylene)-2-methyl-6-(\emph{p}-dimethlaminostyryl)-4H-pyran]
dye produces $\sim$~300~mW of 669-nm light. The dye laser is also
pumped by an Ar ion laser (Coherent Innova 300) operating at 6~W of
power.

Since the population transfer of a weakly collimated atomic beam due
to narrow-band cw lasers is inefficient, an adiabatic passage
technique is utilized to transfer atoms to level B. A detailed
description of this technique and references to earlier work are
given in Ref.~\cite{Nguyen2000}. Briefly, cylindrical lenses are
used to diverge the two laser beams in the interaction region such
that the divergences of the light beams match the atomic-beam
divergence. Due to the Doppler effect, the atoms ``see" a frequency
chirp in laser detuning which adiabatically transfers the population
to the excited state.

The rf electric field is formed between two parallel, 10~cm by 5~cm,
wire grids made of 0.002~in Be-Cu wire. The separation between
adjacent wires is 2.5 mm which minimizes surface area and hence
surface-charge accumulation. The separation between the grids is
$\sim 2.5$~cm. The entire assembly is enclosed inside a
high-permeability magnetic shield which reduces the magnetic field
in the interaction region to $\sim 1$~mG. The 564-nm fluorescence is
collected by a lucite light pipe and detected with a photomultiplier
tube (PMT) with bandpass interference and short-pass filters at the
input window.

\begin{figure}
\includegraphics{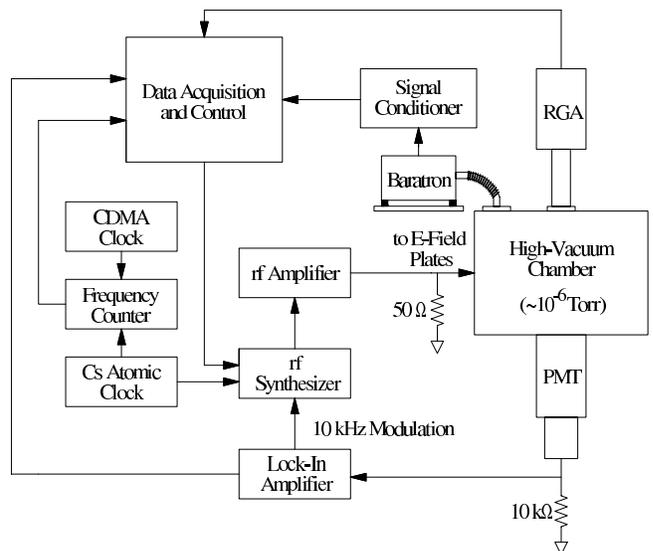}
\caption{\label{fig:setup} Schematic diagram of the rf-generation,
control, and pressure-measurement systems. The rf synthesizer is
referenced to a Cs standard, which is compared to a CDMA (Code
Division Multiple Access) primary reference source. Both clocks have
a long term stability of $\sim 10^{-11}$. The output of the
synthesizer is amplified and then sent to the electric-field plates
and a 50~$\Omega$ power-resistor termination in series with the
larger impedance of the plates. The output of the photomultiplier
tube is sent to the lock-in amplifier for detection and
acquisition.}
\end{figure}

The rf-generation and detection system is shown in
Fig.~\ref{fig:setup}. The frequency-modulated rf field is generated
by a synthesizer (Hewlett Packard 8647A) and sent through a 45-dB
amplifier (ENI 604L). The rf power at the output is $\sim 0.5$~W.
The synthesizer is referenced to a Cs frequency standard (Hewlett
Packard 5061A) with a long term stability ($>\!\!100$~s) of
10$^{-11}$. This frequency standard is compared to a second clock
(Symmetricom TS2700) to monitor its stability. A Signal Recovery
7265 lock-in amplifier provides a 10-kHz modulation signal to the rf
synthesizer and demodulates the signal from the PMT. The signal from
the lock-in along with the clock, pressure, and temperature data are
sent to a computer for data acquisition.

The pressure measurement is provided by a residual gas analyzer (SRS
RGA 200) and a capacitance manometer (MKS Baratron 690A.01). The
residual gas analyzer (RGA) is used to monitor partial pressures of
the background gasses. However, due to unknown gas-species-dependent
sensitivity factors, it is not well suited to measuring the absolute
pressure of various gases. The manometer, on the other hand, is
capable of absolute pressure measurement in a wide range of
pressures from 0.1 Torr to a few $\mu$Torr with a resolution of
10$^{-7}$ Torr. Measurements have shown that the manometer response
is nonlinear below 10~$\mu$Torr. Thus, the total manometer pressure
was limited to 20~$\mu$Torr during the collisional-perturbation
measurements. The manometer signal is processed by a signal
conditioner (MKS 670B) which includes an active heater control
mechanism to stabilize the operating temperature. To further reduce
noise, the manometer is mechanically isolated by a
vibration-isolation mount at the base and by the use of flexible
bellows to connect to the vacuum chamber. The measured statistical
uncertainty in the pressure range used during measurements is $\sim
1$~$\mu$Torr for 100 seconds of integration.

\subsection{\label{sec:proc}Procedure}

\begin{figure}
\includegraphics{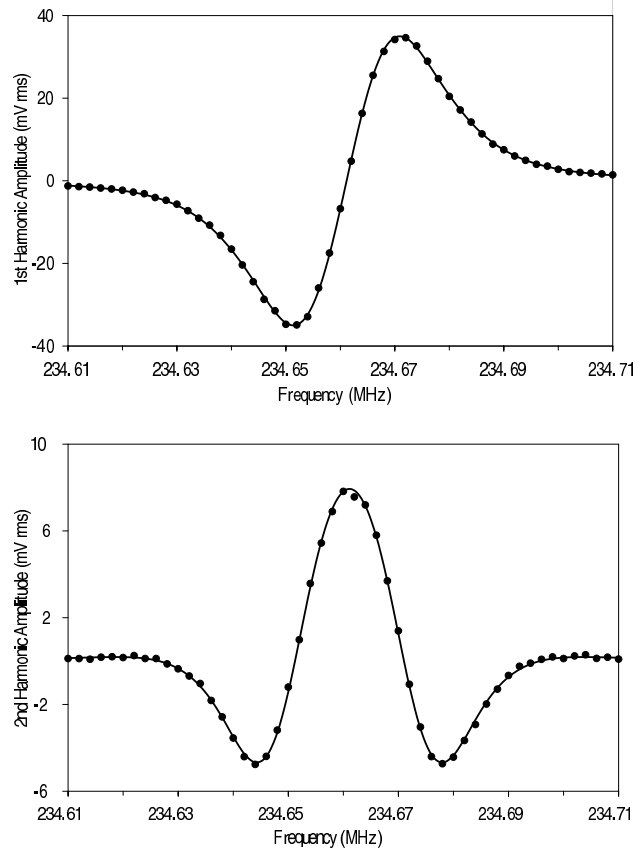}
\caption{\label{fig:h1h2} First- and second-harmonic signals for
10-kHz frequency modulation. Solid line is a fit to the function
derived from first-order time-dependent perturbation theory.}
\end{figure}

To minimize fluctuations such as those due to laser-power drifts or
density fluctuations of the atomic beam, the rf voltage is frequency
modulated at 10~kHz with an equal modulation depth (modulation index
of 1). Since the rf-transition linewidth is $\sim 20$~kHz, this
modulation provides a fast sweep across the absorption line shape.
The outputs from the lock-in amplifier at the first and second
harmonic of the modulation frequency are shown on
Fig.~\ref{fig:h1h2}. The first-harmonic signal is an odd function of
detuning while the second-harmonic signal is an even function. It is
well known that for small modulation amplitudes, the output signal
at the harmonics of the modulation frequency are proportional to the
corresponding derivatives of the absorption line
shape~\cite{Demtroder}. In our system, since the modulation
amplitude is comparable to the rf-transition linewidth, the harmonic
signals contain higher derivative contributions. Therefore, a
fitting function was derived from first-order time-dependent
perturbation theory. This model does not include any effects due to
power broadening or beam velocity distribution and is used to
confirm qualitative behavior of the line shape.

To measure the transition frequency, the ratio of the first- and
second-harmonic signals is used in order to reduce drifts further.
The transition frequency is extracted from the ratio by a two-step
process. First, the lock-in amplitude at the first and second
harmonics and their ratio are recorded
simultaneously~\cite{note:lockin} as the carrier frequency is
scanned 4 kHz around the resonance. For such a small scan compared
to the linewidth, the first harmonic is linear while the second
harmonic is nearly a constant (Fig.~\ref{fig:h1h2}). By fitting a
linear function to the ratio, a conversion parameter from the ratio
to frequency is obtained. Thus, these scans are called calibration
files. Once the calibration is established, the rf carrier frequency
is locked at a fixed value near resonance (3,074,000~Hz and
234,661,000~Hz for the two transitions), and the ratio is measured
repeatedly. These measurements are saved in the so-called
fixed-frequency files. The average of ratios from these measurements
are converted to frequency using the calibration file.

It should be mentioned that before recording a calibration file, a
specific phasing procedure is used to maximize the lock-in-amplifier
signal at both harmonics. Since Signal Recovery 7265 lock-in
amplifier allows for separate phase adjustments for the two
harmonics, the first-harmonic signal phase is adjusted to maximize
the signal at 10~kHz above the fixed-frequency value, while the
second-harmonic signal is maximized at the fixed frequency value.
This procedure is used since the first- and second-harmonic signals
are maximal near these frequencies, respectively. In addition, to
minimize any instability or loss of signal due to laser detuning,
each laser is retuned between recording data files by scanning over
the laser resonances and setting the laser frequency to the
resonance peak. Using this technique for frequency measurements, we
have been able to achieve $\sim$~2- and 6-Hz statistical sensitivity
for 1 second of integration time for the 235- and 3.1-MHz
transitions, respectively.

For collisional-perturbation measurements, foreign gas is introduced
into the chamber through a leak valve (Varian Model 1000). In order
to minimize contamination, the hose connecting the gas cylinder to
the leak valve is evacuated with a separate pump and flushed with
gas several times prior to opening the valve. Then, the leak valve
is opened  and foreign gas is introduced continuously until a
dynamic equilibrium is reached at a manometer pressure reading of
$\sim$~80~$\mu$Torr. Meanwhile, the RGA is monitored for
contamination due to other residual gases in the hose. After the
background gas pressures drop below 10 $\mu$Torr, measurements are
taken at descending pressures between 80 and 20 $\mu$Torr. At each
pressure, one calibration and three fixed-frequency files are taken
for each transition. To check for systematic drifts and signal loss,
two more measurements at intermediate pressures are taken after the
lowest-pressure measurement.

Since foreign gas is introduced from the top of the chamber while
the chamber is continuously pumped from the bottom, a pressure
gradient between the top and the interaction region might exist. Due
to the large conductance of our chamber, we estimate a pressure
gradient of $\sim$~5\%. This estimate is in agreement with the upper
bound extracted from measurements taken with varying pumping speeds.
Since the uncertainty in this pressure calibration error is not well
known, this correction is not included in the results presented in
Sec~\ref{sec:analysis} and Sec.~\ref{sec:dis} in order to preserve
the relative precision of the measurements.

\begin{figure}
\includegraphics{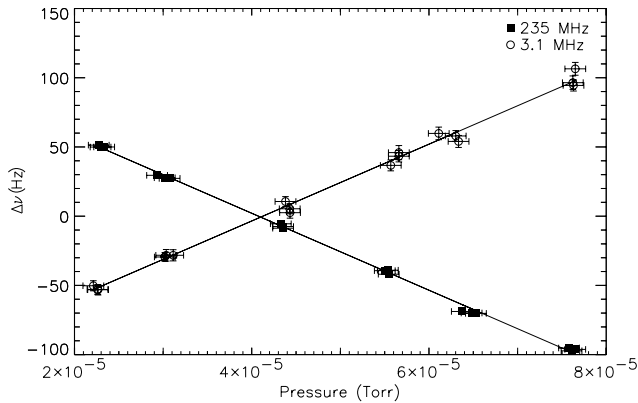}
\caption{\label{fig:shift} Shift of the resonance frequency for both
transitions due to the presence of Krypton. $\Delta \nu$ is the
difference between the transition frequency and the fixed rf
synthesizer carrier frequency (3,074,000 Hz and 234,661,000 Hz for
the two transitions). The solid lines are the least-squares linear
fits to the data.}
\end{figure}

\section{Analysis and Results\label{sec:analysis}}
\subsection{Frequency shifts}
For all the gases except H$_2$ and Ne, which do not shift the
transition frequency, $\sim$~100~Hz shifts were observed over the 10
to 80~$\mu$Torr pressure range. Since these transitions are between
two excited states, the measured shift rates are due to the
difference between the shift of the levels involved in the
transition. As shown in Fig.~\ref{fig:shift}, the shifts for the two
rf transitions are in opposite directions. This is expected since
for the 235-MHz transition, level B is lower in energy than level A,
while for the 3.1-MHz transition, the opposite is the case.
Moreover, since electronic structures of the isotopes are identical,
the absolute value of the shift coefficients should be identical,
which is indeed the case within the precision of the experiment. For
the gases that produce shifts, all but He, shift a given transition
in the same direction.

\begin{table}
\caption{\label{table:shifts} Collisional-shift rates for the
235-MHz and 3.1-MHz transitions in the presence of various gases.
The shift rates are expected to be equal and opposite for the two
transitions since there is a change in sign between the energy
difference of the levels involved. The O$_2$ coefficient was
extracted using a measurement when air was let into the vacuum
chamber. This was only performed for the 235-MHz transition. The
results have been corrected for the presence of the background gases
and the assigned errors include the effects of Dy self-shifts
discussed in Sec.~\ref{sec:syssb}. As discussed at the end of
Sec.~\ref{sec:proc}, the analysis does not include the estimated 5\%
correction due to pressure calibration error.}

\begin{ruledtabular}
\begin{tabular}{lcc}
   & 235-MHz Transition & 3.1-MHz Transition \\
  & shift rates (Hz/$\mu$Torr) & shift rates (Hz/$\mu$Torr) \\
   \hline
  H$_2$ & -0.02 $\pm$ 0.02 & -0.09 $\pm$ 0.06 \\
  N$_2$ & -1.71 $\pm$ 0.03 & 1.72 $\pm$ 0.05 \\
  O$_2$ & -1.97 $\pm$ 0.25 & - \\
  He & 1.25 $\pm$ 0.02 & -1.27 $\pm$ 0.05 \\
  Ne & -0.01 $\pm$ 0.02 & -0.02 $\pm$ 0.05 \\
  Ar & -2.21 $\pm$ 0.05 & 2.14 $\pm$ 0.07 \\
  Kr & -2.78 $\pm$ 0.05 & 2.78 $\pm$ 0.07 \\
  Xe & -2.74 $\pm$ 0.04 & 2.75 $\pm$ 0.07 \\
\end{tabular}
\end{ruledtabular}
\end{table}

The analysis of the shift data is complicated by the effects of
background gases. Since the normal operating pressure of the vacuum
chamber with the oven turned on is $\sim$~5 - 10 $\mu$Torr, the
background gases, dominated mostly by H$_2$ and N$_2$, and to a
lesser extent by O$_2$, make up a significant fraction of the
overall pressure. Moreover, in some cases these gas pressures are
not constant throughout a measurement~\cite{note:procedure} and
contribute to the observed shift rates.

Since the capacitance manometer measures the total gas pressure in
the chamber, RGA partial gas pressures are needed to establish the
dominant-gas pressure and correct for the systematic shifts induced
by the background gases. However, RGA partial pressures must be
corrected with gas-species-dependent sensitivity factors that are
not well determined. We have tried to extract these sensitivity
factors using both data taken during collisional-perturbation
measurements, where the oven is operating at $\sim 1500$~K, and
auxiliary cold-oven measurements, where gas is introduced into the
chamber with the oven turned off. In the latter case, the large
H$_2$ background is absent, and the operating pressure of the vacuum
chamber (without the foreign gas) is $\sim 1$~$\mu$Torr, dominated
by N$_2$. Therefore, the background gases can be ignored and the
sensitivity factors extracted using linear fits to the manometer
versus RGA data.

Extracting these parameters from the hot-oven data relies on the
fact that during measurements where H$_2$ is introduced into the
chamber, the N$_2$ and O$_2$ pressures are stable (we ignore all
other gases including H$_2$O since their pressures are below 5
$\times$ 10$^{-7}$ Torr throughout these measurements). Using this
fact, the H$_2$ sensitivity factor, $\beta_{H2}$, was extracted in
the same way as the auxiliary cold-oven method, ignoring the other
gases. The N$_2$ sensitivity factor, $\beta_{N2}$, was extracted
from measurements where N$_2$ was introduced into the chamber, but
by accounting for the background H$_2$ pressure (using the extracted
$\beta_{H2}$). Finally, the O$_2$ sensitivity factor, $\beta_{O2}$,
was extracted taking both the N$_2$ and H$_2$ pressures into account
using data where air was introduced into the chamber.

The RGA sensitivity factors extracted using the cold-oven and
hot-oven data disagree at $\sim$~10 - 20~\% level, suggesting that
they are also dependent on the background-gas pressures. As a
result, in the rest of the analysis, sensitivity factors extracted
from in-situ hot-oven measurements were used with larger error bars
to accommodate for the results obtained by the auxiliary cold-oven
measurements. In addition, in order to minimize the uncertainty
introduced by the use of RGA sensitivity factors, the dominant gas
pressures, $P_{gas}$, were calculated using
\begin{equation}
P_{gas}=P(CM)-\sum_i \beta_i P_i(RGA),
\end{equation}
where $P(CM)$ is the manometer pressure, $P_i(RGA)$ and $\beta_i$
are the RGA partial pressure and sensitivity factor of a background
gas species $i$, respectively. Since the background-gas pressures
are at least two times smaller than the dominant-gas pressure, by
using this method, the uncertainty due to the sensitivity factors
are reduced to levels comparable to or smaller than the manometer
uncertainty of 1~$\mu$Torr.

Once the sensitivity factors of the background gases were
determined, an iterative approach was used to extract the Dy
rf-transition-shift rates for H$_2$, N$_2$, and O$_2$. The iterative
approach is necessary for these gases because for a measurement
where one of these gases dominates, the other two are background
gases. As a result, to extract the correct shift rate for one of
them requires the knowledge of the shift rates of the other two,
which, in turn, are dependent on the shift rate that is being
extracted. So, on the first iteration, the H$_2$ shift rate was
extracted by taking into account the background N$_2$ and O$_2$
pressures but not the shifts due to them. Then, the N$_2$ shift rate
was extracted similarly, by taking into account both the H$_2$ and
O$_2$ pressures, as well as shifts due to H$_2$. Finally, the O$_2$
shift rate was extracted by taking into account both N$_2$ and H$_2$
pressures and shifts (only for the 235-MHz transition since no air
data were taken for the 3.1-MHz transition). Then, these first-order
coefficients were used to extract higher-order coefficients where
pressures and shifts due to all the background gases were used in
every step. This iterative process converged within three
iterations, improving $\chi^2$ at every step.

The rest of the gases were easier to analyze since they do not
appear as background gases in other measurements. The coefficients
for these gases were determined by taking into account the
background H$_2$, N$_2$ and O$_2$. For the 3.1-MHz transition, the
O$_2$ shift rate for the 235-MHz transition with the opposite sign
was used. The results are presented in Table~\ref{table:shifts}.

\subsection{Broadening}

\begin{figure}
\includegraphics{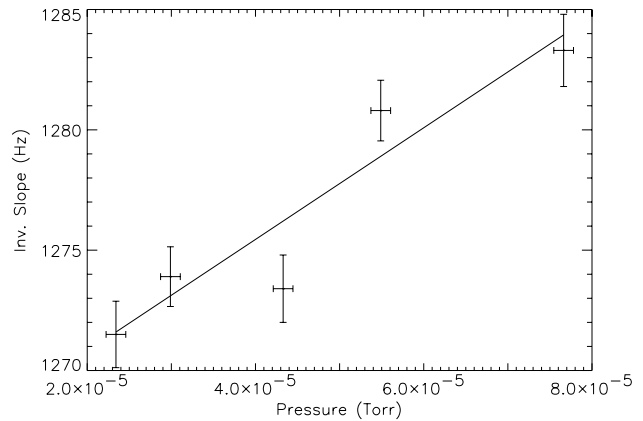}
\caption{\label{fig:broaden} Collisional broadening for the 235-MHz
transition due to Kr without the phasing-effect correction. On the y
axis, the inverse of the normalized first harmonic slope, which is
directly proportional to the linewidth is plotted. The solid line is
the least-squares linear fit to the data.}
\end{figure}

Since the main focus of this study is frequency shifts, the
experimental technique we utilize is optimized for frequency
measurements. Nevertheless, the broadening rates can also be
extracted from the calibration files. However, determination of the
broadening rates is more difficult than the shift rates since, as
discussed below, the sensitivity of the calibration data to
linewidth is low. In addition, there is a systematic effect caused
by the phasing procedure used (see Sec.~\ref{sec:proc}) coupled with
the collisional shift of the resonance frequency that results in an
enhancement of apparent broadening rates for one transition and
reduction for the other.

The slope of the first-harmonic signal near resonance, normalized by
the second harmonic can be extracted by fitting a linear function to
the ratio from the calibration file data, as described in
Sec.~\ref{sec:proc}. For small modulation amplitudes, the
first-harmonic signal is proportional to the first derivative of the
absorption line shape and the second harmonic to the second
derivative. Near resonance, to first order in frequency detuning,
both of these functions have a 1/$\gamma^3$ dependence. As a result,
the slope of the ratio is insensitive to the linewidth. However, in
our system the modulation amplitude is comparable to the linewidth,
and the harmonics contain higher-order derivatives which result in a
small dependence of the slope on the linewidth. In practice, the
inverse of this slope is the useful quantity in converting the ratio
to frequency. It is proportional to the linewidth for small changes
in the linewidth~\cite{note:broadening} and increases with gas
pressure due pressure broadening~(Fig.~\ref{fig:broaden}).

\begin{table}
\caption{\label{table:broaden} Collisional broadening rates for the
235-MHz transition in the presence of various gases. The large
uncertainties are due to statistical noise and the uncertainty in
the conversion factor from the inverse slope to linewidth. The O$_2$
result is presented as an upper bound at 2$\sigma$ confidence level.
The results do not include the estimated 5\% uncertainty due to
pressure gradients in the chamber.}
\begin{ruledtabular}
\begin{tabular}{lc}
   & 235-MHz Transition\\
 & broadening rates (Hz/$\mu$Torr) \\
   \hline
  H$_2$ & 8.6 $\pm$ 1.9 \\
  N$_2$ & 3.0 $\pm$ 1.7 \\
  O$_2$ & $<$ 8.6 \\
  He & 5.3 $\pm$ 1.9 \\
  Ne & 1.4 $\pm$ 1.3 \\
  Ar & 4.9 $\pm$ 1.4 \\
  Kr & 5.3 $\pm$ 1.4 \\
  Xe & 2.8 $\pm$ 2.0  \\
\end{tabular}
\end{ruledtabular}
\end{table}

The conversion factor from the inverse slope to linewidth is
extracted from measurements where calibration files and line-shape
scans were taken simultaneously for different rf input powers, which
lead to power broadening. The extracted conversion factor of 31(5)
agrees with the conversion factor of 28.2 obtained from the
first-order perturbation theory calculation by linearizing the ratio
near resonance. In the analysis of the data, the experimental value
is used and errors propagated accordingly.

\begin{figure}
\includegraphics{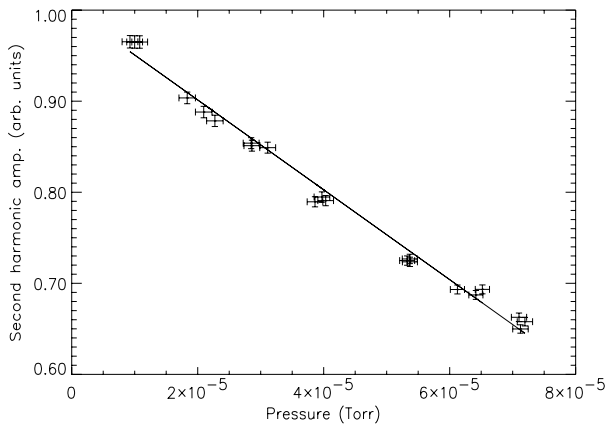}
\caption{\label{fig:quenching} Decrease in the second harmonic
signal for the 235-MHz transition due to Xe . The signal has been
normalized to one at zero pressure. The solid line is the
least-squares fit to the data.}
\end{figure}

There is a systematic effect related to the phasing procedure which
is due to the dependence of the first-harmonic slope near resonance,
and hence the slope of the ratio, on the frequency at which the
phase is adjusted to maximize the signal (which we define as phasing
frequency). For small changes in the phasing frequency, the
experimental measurements show a linear dependence with 0.025(3) and
0.029(5) Hz change in the inverse slope per Hz change in the phasing
frequency for the 235- and 3.1-MHz transitions, respectively. For
all the gases where shifts are observed, with the exception of He,
which shifts the transition frequency in the opposite direction,
this effect leads to an enhancement of the broadening rates for the
235-MHz transition and a reduction for the 3.1-MHz transition. This
is due to the fact that as the gas pressure is changed, the
transition frequency shifts but the phasing frequency is fixed at
10~kHz above the carrier frequency used for the fixed-frequency
measurements. As a result, the phasing frequency changes with
respect to the transition frequency. For example, from the Kr shift
data shown in Fig.~\ref{fig:shift}, at low pressures, the rf carrier
frequency for the 235-MHz transition is 50~Hz below the transition
frequency, while at high pressures, it is 100~Hz above the
transition frequency (the y axis of the figure is transition
frequency minus carrier frequency). As a result, while at low
pressures, the phasing frequency with respect to the transition
frequency is less than 10~kHz, at high pressures, it is greater than
10~kHz. This leads to an enhancement of apparent broadening since at
low pressures the observed inverse slope is less than the value that
would be observed without this effect, and the opposite is the case
at high pressures. The 3.1-MHz transition is shifted in the opposite
direction, and as a result, the observed broadening is reduced for
this transition. This is a significant effect since the changes in
the transition frequency lead to 3 - 4 Hz changes in the inverse
slope. In comparison, as seen in Fig.~\ref{fig:broaden}, the changes
in the inverse slope due to the gases are 10 - 15 Hz.

\begin{table*}
\caption{\label{table:quenching} Signal-attenuation rates for the
235-MHz and 3.1-MHz transitions and upper bounds for collisional
quenching cross sections in the presence of various gases. The
results have been corrected for the presence of the background gases
and the assigned error bars include systematics due to beam-density
variations.The results do not include the estimated 5\% uncertainty
due to pressure gradients in the chamber. For the combined cross
sections, gas temperature of 300(5)~K and Dy beam temperature of
1500(50)~K  are assumed.}
\begin{ruledtabular}
\begin{tabular}{cccc}
   & 235-MHz Transition & 3.1-MHz Transition & Quenching \\
   & $\epsilon$ ($10^{-3}\  1/\mu$Torr)& $\epsilon$ ($10^{-3}\  1/\mu$Torr) & $\avg{\sigma_q v}$ ($10^{-10}$ cm$^3$/s) \\
   \hline
  H$_2$  & $4.1\pm0.2$ &  $3.9\pm0.2$ & $< 13$\\
  N$_2$  & $4.4\pm0.5$ &  $4.3\pm0.7$ & $< 14$\\
  O$_2$  & $10\pm1$&  - & $< 32$\\
  He &  $2.6\pm0.2$ &  $2.6\pm0.2$ &$< 8$\\
  Ne &  $2.3\pm0.2$ &  $2.6\pm0.2$ &$< 7.6$\\
  Ar &  $3.3\pm0.3$ &  $3.7\pm0.2$ & $< 11$\\
  Kr &  $3.9\pm0.2$ &  $4.0\pm0.2$ & $< 13$\\
  Xe &  $4.9\pm0.2$ &  $5.2\pm0.2$  & $< 16$\\
\end{tabular}
\end{ruledtabular}
\end{table*}

The results that have been corrected for this systematic effect, as
well as the systematic effect due to the presence of the background
gases, are presented in Table~\ref{table:broaden}. Even with this
correction, the statistical noise coupled with the uncertainty in
the conversion from inverse slope to linewidth lead to poor fits for
the 3.1-MHz data. Consequently, the broadening results for this
transition are not presented. For the same reasons, only an upper
bound for the 235-MHz O$_2$ broadening rate is presented. Due to
small O$_2$ pressure variation ($\leq$~1.1~$\mu$Torr), the omission
of the O$_2$ correction is not problematic in this case since even
if the broadening rate for O$_2$ were 8.6~Hz/$\mu$Torr (the upper
2$\sigma$ bound from Table~\ref{table:broaden}), this would lead to
an error of $\lesssim$ 5\% for the measurements that exhibit the
largest O$_2$ pressure variation.

\subsection{\label{sec:analq}Signal Attenuation}
During the fixed-frequency measurements, $\sim$~15 - 40~\% decrease
in the second-harmonic-signal amplitude, which is nonzero at
resonance, as a function of foreign-gas pressure was observed
(Fig.~\ref{fig:quenching}). This signal attenuation can be
attributed to either decrease in the beam density due to elastic
scattering or decrease in the fluorescence intensity due to
inelastic quenching. The signal amplitude decreases with the
distance the beam travels through the foreign gas:
\begin{equation}
<\!\!S(v)\!\!>=G I_0<\!\!e^{-l/\lambda(v)}\!\!>,
\end{equation}
where $G$ is the efficiency of the detection system, $I_0$ is the
fluorescence intensity at zero pressure, $\lambda$ is the mean free
path, and $l$ is the beam path length through the foreign gas. Here,
$<...>$ denotes average over the beam velocity distribution. Since
$\lambda$ is inversely proportional to pressure, the signal for
small pressures can be written as:
\begin{equation}
\label{eq:epsilon}\frac{<\!\!S(v)\!\!>}{S_0}=<\!\!e^{-l/\lambda(v)}\!\!>\approx
<\!\!1-l/\lambda(v)\!\!>=(1-\epsilon p),
\end{equation}
where $S_0$ is the second-harmonic signal at zero pressure, and
$\epsilon$ is the attenuation rate.

The analysis of the quenching data is similar to the analysis of the
shift and broadening. First, the iterative process was used for the
dominant gases in the background, H$_2$, N$_2$, and O$_2$. Then, the
values extracted were used in the analysis of the remaining gases.
The attenuation rates tabulated in Table~\ref{table:quenching}
include systematic corrections due to clogging of the unheated
photon collimator (Fig.~\ref{fig:setup}) and temperature drift of
the ovens. To detect any changes in the signal amplitude due to
clogging, the two measurements at intermediate pressures taken after
the lowest pressure points, as discussed in Sec.~\ref{sec:proc},
were used. The oven temperature was monitored by a thermocouple and
changes in the beam density due to temperature fluctuations were
estimated using Dy saturated vapor-pressure dependence on
temperature (see Sec.~\ref{sec:syssb}.)

The interpretation of the signal attenuation data is complicated
since it is difficult to separate elastic scattering events that
decrease beam density from inelastic quenching that decreases the
population of level B. Nevertheless, we expect the elastic
scattering contribution to be small since although typical elastic
cross sections are $\sim$~10$^{-14}$~cm$^2$ (corresponding to a mean
free path of $\sim$~30 cm at 100~$\mu$Torr), large angle scattering
necessary for significant collimation loss is suppressed by more
than one order of magnitude due to the fact that differential cross
section is sharply peaked at small scattering angles~\cite{Ramsey}.
Using the results derived in Ref.~\cite{Massey} for hard-ball
scattering, we have estimated that the differential cross section
becomes appreciable below 0.1~-~0.01~radian in the center-of-mass
frame. For a head-on collision, this results in 6~-~7~mrad deviation
for the Dy beam in the laboratory frame. This deviation is small
compared to the collimation half angle of 0.1~radian used in our
experiment. However, since this estimate uses a hard-ball model,
which depends on few assumptions such as a specific minimum distance
between the centers of the spheres, as opposed to more complicated
interactions such as quasi-molecular formation, it can vary by a
multiplicative factor of 2 - 5. As a result, we only present upper
bounds for the inelastic quenching cross sections.

The attenuation rates can be related to the quenching cross sections
through a characteristic mean free path, which can be expressed as
\begin{equation}
\label{eq:mfp} \widetilde{\lambda}=\frac{\avg{v_B}}{n_G\avg{\sigma_q
v}}
\end{equation}
where $\avg{v_B}=3/4(2\pi\, k\, T_B/m_B)^{1/2}$ is the average beam
velocity given by a modified Maxwellian distribution, $n_G$ is the
density of foreign gas, $v$ is the relative velocity of the gas and
beam particles, and $\avg{\sigma_q v}$ denotes an average over both
the gas and the beam velocity distributions.

Using Eq.~(\ref{eq:mfp}) and Eq.~(\ref{eq:epsilon}), the quantity
$\avg{\sigma_q v}$ can be written in terms of $\epsilon$  as
\begin{equation}
\avg{\sigma_q v}=\frac{3}{4}\sqrt{\frac{2\pi
k}{m_{\text{Dy}}}}\,\frac{k\,\epsilon}{l}\,T_G\, T^{1/2},
\end{equation}
where $T_G$ and $T$ are the temperatures of the foreign gas and the
oven, respectively. The path length, $l$, is not well defined in our
experimental setup. Since the light pipe has a width of 6.7~cm, it
collects fluorescence from atoms at various path lengths. The
tabulated results assume a path length of 5.5~cm, which is the
distance between the laser excitation region and the center of the
light pipe. The results are summarized in Table
\ref{table:quenching}.

\section{\label{sec:syssb}Systematic Uncertainties}

In addition to systematic effects considered above, Dy*(excited Dy
atoms)-Dy* and Dy*-Dy collisions within the beam can lead to a
systematic change in the transition frequency and linewidth, if the
beam density does not remain constant during the collisional
measurements. We have performed independent measurements of
self-collisions in our system and present here an estimate of their
contribution to the systematic uncertainty.

The beam density can be controlled by varying the oven temperature.
The temperature dependence of Dy saturated-vapor pressure is given
by the following empirical formula~\cite{CRC}:

\begin{equation}
\label{eq:vapor}
 P=10^{12.460-15336/T-1.114\log(T)},
\end{equation}
where $P$ is the pressure in Torr and $T$ is the temperature in
Kelvin.

\begin{figure}
\includegraphics{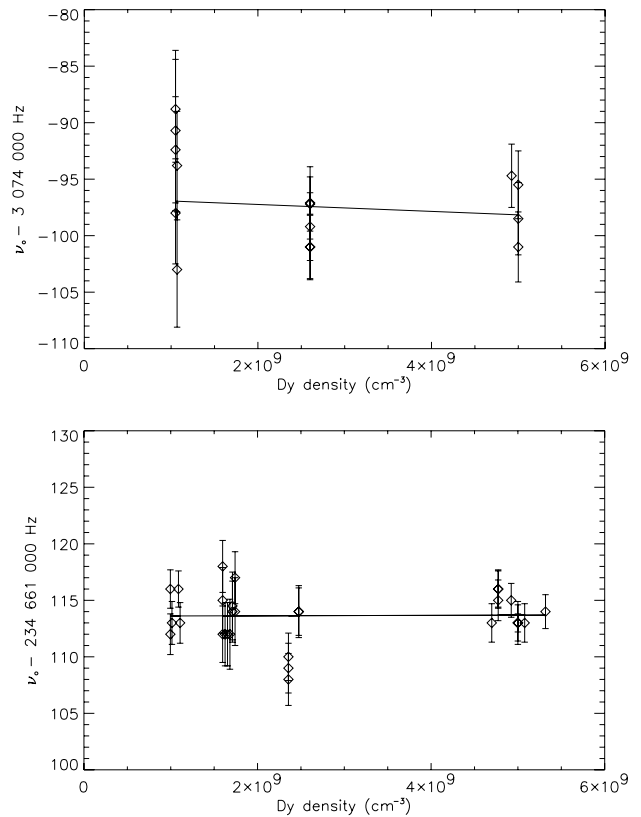}
\caption{\label{fig:self} Shifts due to collision of the Dy atoms in
the beam. Beam density was varied by changing the oven temperature
from 1170$^\circ$C to 1077$^\circ$C which correspond to the
estimated atomic densities in the interaction region shown in the
plots.}
\end{figure}

We have varied the oven temperature, which was monitored by a
thermocouple, from 1077$^\circ$C to 1170$^\circ$C, while monitoring
the frequency for both transitions. This temperature range
corresponds to a Dy density of ($0.3~-~
1.5$)~$\times10^{14}$~cm$^{-3}$ in the oven. Earlier absorption
measurements~\cite{Nguyen1997} have shown that the density of atoms
in the population and detection region is $\sim$~5
$\times$~10$^{9}$~cm$^{-3}$ at the normal operating temperature of
1170$^\circ$C. This density is in agreement with estimates based on
effusive flow. To obtain the Dy density in the interaction region
for various temperatures, the scaling factor between the calculated
density in the oven at 1170$^\circ$C and the absorption measurements
was applied uniformly to all the calculated densities. The extracted
self-shift rates were 0.2(2.0)~$\times$~10$^{-10}$~Hz~cm$^3$ and
-3.1(6.0)~$\times$~10$^{-10}$~Hz~cm$^3$ for the 235-MHz and 3.1-MHz
transitions, respectively (Fig.~\ref{fig:self}). An upper bound for
the magnitude of the self-shift rate can be estimated by noting that
the rates, $m$, for these two transitions should be the same in
magnitude but opposite in sign:
\begin{eqnarray}
|m|=\frac{1}{2}|m_{235}-m_{3.1}|,\\
0=|m_{235}+m_{3.1}|, \nonumber
\end{eqnarray}
where the second equation provides a way to estimate errors. The
resulting upper bound on the magnitude of the self-shift rate is
$<$~4.9~$\times$~10$^{-10}$~Hz~cm$^3$ at 1$\sigma$ confidence level.
Since changing the oven temperature does not discriminate between
Dy* and Dy densities, this upper bound is an effective rate due to
both Dy*-Dy* and Dy*-Dy collisions. A simple way to address this
ambiguity would be to change the laser powers simultaneously to
change the fraction of excited Dy atoms. This method may be pursued
in a future study.

Using the extracted upper bound, we can estimate the systematic
contribution of self-collisions to the measured shift and broadening
rates. This requires a mechanism that changes Dy beam density as a
function of foreign-gas pressure. As a worst-case scenario, we
assume that all of the observed signal attenuation is due to the
decrease in the Dy density as a result of scattering with the
foreign-gas atoms. Depending on the gas, this results in a change of
$\sim$~($0.9 - 2.2$)~$\times$~10$^9$~cm$^{-3}$ in the Dy density,
which corresponds to $\sim~0.4 - 1.1$~Hz change in the magnitude of
the transition frequency from low to high has pressures. The
correction to the foreign-gas shift rates are in the range of $0.005
- 0.01$~Hz/$\mu$Torr. These uncertainties have been included in
Table~\ref{table:shifts}.

\begin{table*}
\caption{\label{table:sbcrosssections} Collisional shift and
broadening cross sections and their ratios at foreign-gas
temperature of 300(5)~K. The shift cross sections for the two
transitions have been combined into a single result with the
following sign convention: a positive frequency shift for the
3.1-MHz transition gives a positive cross section.}
\begin{ruledtabular}
\begin{tabular}{cccc}
  & $\avg{\sigma_s v}$ ($10^{-10}$ cm$^3$/s) & $\avg{\sigma_b v}$ ($10^{-10}$ cm$^3$/s) & $\!\!\avg{\sigma_s v}\!\!/2\avg{\sigma_b v}$ \\
  \hline
  H$_2$ & $0.05\pm0.04$ & $8.4\pm1.9$ & $0.003\pm0.002$  \\
  N$_2$ & $3.34\pm0.07$ & $2.9\pm1.7$ & $0.6\pm0.3$  \\
  O$_2$ & $3.9\pm0.5$ & $<$ 8.4 & $<$ 0.2  \\
  He & $- 2.44\pm0.05$ & $5.2\pm1.9$ & $- 0.2\pm0.1$  \\
  Ne & $0.02\pm0.04$ & $1.4\pm1.3$ & $0.01\pm0.02$ \\
  Ar & $4.3\pm0.1$ & $4.8\pm1.4$ & $0.4\pm0.1$  \\
  Kr & $5.4\pm0.1$ & $5.2\pm1.4$ & $0.5\pm0.1$ \\
  Xe & $5.3\pm0.1$ & $2.7\pm2.0$ & $1.0\pm0.7$   \\
\end{tabular}
\end{ruledtabular}
\end{table*}

Since, in general, broadening should be of the same order of
magnitude as shifts, the uncertainty due to self-broadening is
negligible compared to the errors quoted on
Table~\ref{table:broaden}.

\section{Discussion\label{sec:dis}}

The adiabatic impact approximation results give a Lorentzian line
shape with the transition frequency and linewidth modified
by~\cite{note:discussion}:
\begin{eqnarray}
\Delta \gamma=\frac{1}{2\pi}(2n\avg{v\sigma_b})=\frac{\avg{v\sigma_b}}{\pi k T}p,\\
\Delta
\nu=\frac{1}{2\pi}(n\avg{v\sigma_s})=\frac{\avg{v\sigma_s}}{2\pi k
T}p,
\end{eqnarray}
where $k$ is the Boltzmann constant, $p$ and $T$ are the foreign-gas
pressure and temperature, $\avg{...}$ denotes an average over both
the gas and the beam velocity distributions. The shift and
broadening cross sections are given by (see, for example,
Ref.~\cite{Sobelman}):
\begin{subequations}
\label{eq:crosssections}
\begin{eqnarray}
\sigma_s=2\pi\int_0^\infty b\sin(\phi(b)) \, db,\\
\sigma_b=2\pi\int_0^\infty b(1-\cos(\phi(b)) \, db.
\end{eqnarray}
\end{subequations}
Here, $\phi(b)$ is the induced phase shift which depends on the form
of the interaction potential. These potentials are usually expressed
as V$_k$=C$_n^k$/r$^n$ for the $k^{th}$ excited state with the van
der Waals (n=6) being the primary long range interaction for neutral
atoms. For this potential, the collisional cross sections for a
transition between arbitrary states a and b are given by
\begin{subequations}
\label{eq:vdWalls}
\begin{eqnarray}
\avg{\sigma_b}=2\pi(0.602)\left(\frac{3\pi |C_6'|}{8\avg{v}}\right)^{2/5},\\
\avg{\sigma_s}=\textrm{sign}(C_6')\ 2\pi(0.438)\left(\frac{3\pi
|C_6'|}{8\avg{v}}\right)^{2/5},
\end{eqnarray}
\end{subequations}
where $C_6'=(C_6^a-C_6^b)/\hbar$.

As illustrated by this example, $\sigma_s$/2$\sigma_b$ has a value
of 0.364, independent of $C_6'$. In fact, within the adiabatic
impact approximation, a unique ratio is predicted for each
interaction potential which can be used to determine the dominant
form of the interaction. Since cross sections can have a dependence
on the relative velocities, the quantities $\avg{\sigma v}$ for both
broadening and shifts extracted from our experimental results are
tabulated in Table~\ref{table:sbcrosssections}, as well as their
ratio~\cite{note:discussion2}. The sign convention for the shift
cross sections has been chosen such that a positive frequency shift
for the 3.1-MHz transition (where level B has higher energy than
level A) gives a positive cross section. The measured ratios vary
widely for different gases. In foreign gas perturbations, this is a
common situation since in addition to long range attractive
interactions, short range repulsive forces can be important. This is
especially true for H$_2$, He and Ne which do not interact strongly
via the van der Waals interaction~\cite{Corney}. Our results are in
agreement with this trend: H$_2$ and Ne cause no shifts, while He
induces an opposite shift as a result of dominant repulsive forces.
On the other hand, all three gases show broadening comparable to the
other gases. This can be understood by observing the integrands of
Eqs.~(\ref{eq:crosssections}). While the integrand for the shift
cross section oscillates around zero, the integrand for the width
cross section is always positive, yielding a nonzero broadening even
for short-range interactions.

\section{Implications for the $\alpha$-variation experiment}
Collisional shifts are one of the most important systematics for an
experiment searching for the temporal variation of $\alpha$ using Dy
since they mimic frequency shifts that would be induced by variation
in $\alpha$. The $\alpha$ dependence of different isotopes and
hyperfine-level manifolds of odd isotpes are the
same~\cite{note:alphadot}. As a result, the entire manifold of level
B shifts identically with respect to the manifold of level A.
Monitoring different transitions where energy of level A is larger
than the energy of B, and vice versa provides a powerful tool in
detecting and eliminating certain systematics since the shift
induced by $\alpha$ variation should have the same magnitude but
opposite sign for these transitions. However, as seen in Table
~\ref{table:shifts}, this behavior is identical to that of the
shifts induced by collisions, and therefore, gas pressures must be
kept stable to 1 part in 10$^{4}$ for a background pressure of
$\sim$~1~$\mu$Torr, or lowered to $\lesssim$~3~$\times$~10$^{-10}$
Torr to achieve the desired mHz-level stability over a year.

This conclusion is in agreement with the estimates provided in
Ref.~\cite{Nguyen2004}. However, it was also argued in
Ref.~\cite{Nguyen2004} that a suppression of collisional shifts
might exist due to the fact that the transitions are nominally
between f and d inner-shell electrons and that inelastic quenching
might remove atoms experiencing collisions from levels A or B so
that they would not contribute to the signal. Our present results do
not reveal such suppression of collisional shifts.

With the current experimental setup, the operating pressure is
$\sim$ 5 $\mu$Torr, mostly dominated by hydrogen and nitrogen. A new
vacuum chamber that can achieve 10$^{-10}$ Torr with a Dy oven
operating at 1500 K is under construction. In the mean time, data
that is being acquired with the current chamber will be corrected
using the measured collisional shift coefficients to achieve an
estimated stability of 2 Hz over a year, which corresponds to
sensitivity of $|\dot{\alpha}/\alpha |\sim 10^{-15}$/yr. This
sensitivity is comparable to the best current laboratory
limits~\cite{Bize, peik}.
\section{Conclusion}
In summary, we have reported collisional shift and broadening rates,
and cross sections for radio-frequency transitions between excited
states of $^{162}$Dy and $^{163}$Dy in the presence of 10 to 80
$\mu$Torr of H$_2$, N$_2$, O$_2$, He, Ne, Ar, Kr, and Xe. In
addition, we have placed upper bounds on quenching cross sections
using the pressure dependence of the signal amplitude. A
frequency-modulation technique used for the search of the temporal
variation of the fine-structure constant was utilized to measure the
small perturbations induced at such low pressures. The extracted
cross sections are similar to cross sections for optical
transitions~\cite{Allard&Kielkopf}.

To improve our understanding of the interaction potentials, several
avenues may be pursued in the future. As seen in
Eqs.~(\ref{eq:vdWalls}), the possible velocity dependence of cross
sections can be exploited by varying the gas temperature. Moreover,
a careful study of the line shape far off resonance would be
informative due to strong temperature dependences and satellite-line
formations at higher pressures~\cite{Corney}.
\begin{acknowledgments}
We thank V. V. Yashchuk, A. Lapierre, D. F. Kimball, D. English, J.
E. Stalkner, and W. Gawlik for valuable discussions. We are grateful
to E. D. Commins for his support and encouragement. This work was
supported in part by the UC Berkeley-LANL CLC program and NIST
Precision Measurement Grant.
\end{acknowledgments}

\bibliography{pracollisions,footnotes}

\begin{thebibliography}{38}
\bibitem[{not({\natexlab{a}})}]{note:lockin}
\bibinfo{note}{Signal Recovery Model 7265 lock-in amplifier is capable of
  providing both harmonics and their ratio simultaneously through digital
  signal processing.}
\bibitem[{not({\natexlab{b}})}]{note:procedure}
\bibinfo{note}{This is mainly true for N$_2$ and O$_2$ due to residual air in
  the hose connecting the gas cylinder to the leak valve that is introduced
  into the chamber with the foreign gas.}
\bibitem[{not({\natexlab{c}})}]{note:broadening}
\bibinfo{note}{The function relating the linewidth to inverse slope is
  nonlinear in general, but for small changes, as in the case of present
  measurements, it can be linearized. As a result of this linearization, in
  addition to the multiplicative conversion factor, there is a large,
  unphysical y-intercept that must be used to convert inverse slopes to
  linewidths. However, this y-intercept is irrelevant for the extraction of the
  broadening rates since it is independent of pressure.}
\bibitem[{not({\natexlab{d}})}]{note:discussion}
\bibinfo{note}{The additional factor of two in the linewidth comes from the
  definition of the Lorentzian line shape which is proportional to
  $1/((\omega-\omega_0)^2+(\gamma/2)^2)$.}
\bibitem[{not({\natexlab{f}})}]{note:alphadot}
\bibinfo{note}{There are differences due to hyperfine interactions, but these
  are suppressed by a factor of $m_e/m_p$.}
\end{thebibliography}
\end{document}